\documentclass[sort&compress,aps,a4paper,superscriptaddress]{revtex4}%NOT showpacs,showkeys
\usepackage{amsmath}
\usepackage[dvips]{graphicx}
\usepackage{epsfig}
\usepackage{color}
\usepackage{enumerate}
\usepackage{bm}
\usepackage{amssymb}
\newcommand{\hz}{\hat{\mathbf{z}}}

\begin{document}
\title{Gyromap for a two-dimensional Hamiltonian fluid model derived from Braginskii's closure for magnetized plasmas}
\author{O. Izacard}
\author{C. Chandre}
\author{E. Tassi}
\affiliation{Centre de Physique Th\'{e}orique - UMR 6207, Campus de Luminy - case 907, 13288 Marseille cedex 09, France}
\author{G. Ciraolo}
\affiliation{Laboratoire de M\'{e}canique, Mod\'{e}lisation \& Proc\'{e}d\'{e}s Propres - UMR 6181, Technop\^{o}le de Ch\^{a}teau-Gombert, 13451 Marseille cedex 20, France}

\begin{abstract}
We consider a plasma described by means of a two-dimensional fluid model across a constant but non-uniform magnetic field $\mathbf{B} = B(x,y) \mathbf{\hat{z}}$. The dynamical evolution of the density and the vorticity takes into account the interchange instability and magnetic field inhomogeneities. First, in order to described the Finite Larmor Radius effects we apply the gyromap to build a Hamiltonian model with ion temperature from a cold-ion model. Second, we show that the gyromap is justified using Braginskii's closure for the stress tensor as well as an apt ordering on the fluctuating quantities.
\end{abstract}
\date{\today}
\maketitle

\section{Introduction}
In laboratory plasma devices like tokamaks or in astrophysical plasmas, two-dimensional reduced fluid models have been used to investigate various instabilities and their impact on turbulent transport. These reduced models are very useful in practice since their theoretical and numerical analyses are simpler than for complete models. At the same time they display important features of the underlying turbulent phenomena with more clarity than the complete models. The standard procedure used for the derivation of these reduced models is to start from the equations of motion of a parent model (e.g., that given by the continuity and the momentum equations) and to perform an ordering with respect to one or several small parameters on the variables of the system (e.g., density and potential fluctuations) and on the electromagnetic field.
This ordering is suggested by the physics and the geometry of the phenomenon under consideration (e.g., one might want to filter out irrelevant time or length scales). 
It is desirable that the reduced model preserves the main ingredients of the parent model under this reduction procedure, as for instance the Hamiltonian symmetry. Typical parent models have indeed this Hamiltonian symmetry, e.g., the collisionless Maxwell-Vlasov equations~\cite{Morrison80PLA,Marsden82} or ideal magnetohydrodynamics~\cite{Morrison80PRL}. However, in such a derivation procedure obtained by working at the level of the equations of motion, neglecting terms can produce ``mutilations''~\cite{Morrison05} of the original fluid equations by breaking this symmetry. These mutilations, which introduce incorrect dissipative terms, have a drastic impact on the properties of the physical system. The loss of the Hamiltonian structure indeed generates a more intricate interpretation of the numerical results, and in general clouds the relation with the parent model and its properties. In fact, as shown for example in models describing magnetic reconnection~\cite{Grasso01,Tassi10}, the Hamiltonian structure helps the interpretation and explanation of complex physical processes observed in numerical simulations. Among the other advantages of the Hamiltonian structure, there is the possibility of identifying systematically invariants of motion (e.g., Casimir invariants or conserved quantities linked to continuous symmetries), and applying tools of Hamiltonian perturbation theory and methods to investigate the stability of equilibria.\\
However, for some reduced models, the important property of possessing a Hamiltonian structure has not been shown. For instance, this is the case of the two-dimensional fluid model taken from Ref.~\cite{Haz85_86}. This four-field model includes Finite Larmor Radius (FLR) effects as in Refs.~\cite{Hsu86,Brizard92}, the drift velocity ordering as in Ref.~\cite{Mikhailovskii71} and the gyroviscous terms as in Refs.~\cite{Hinton71,Scott03}. Moreover, magnetic inhomogeneities are kept in the continuity equation as in Refs.~\cite{Haz85_86,Haz87,Eic96,Gar01}. For this particular model, the Hamiltonian structure has not been found even though the model has been shown to be energy conserving. In this paper, we consider the main obstacle associated with the search for the Hamiltonian structure of this four-field model which is already encapsulated in a two-field model for the density and the vorticity fields. This two-field model is obtained from the four-field model by suppressing the parallel dynamics (along the magnetic field), and the poloidal magnetic and parallel flow fluctuations, and reads
\begin{eqnarray}
\displaystyle
\frac{\partial n}{\partial t} &=& - \left[ \phi , n \right] - \left[ \phi , \frac{1}{B} \right] + \left[ n , \frac{1}{B} \right],\label{eq:TOKAM:continuity} \\
\nonumber\\
\displaystyle
\frac{\partial \Delta\phi}{\partial t} &=& - \left[ \phi , \Delta\phi \right] + (1+T) \left[ n , \frac{1}{B} \right] - T \bm{\nabla} \cdot \left[ n , \bm{\nabla}\phi \right],
\label{eq:TOKAM:vorticity}
\end{eqnarray}
where $n=n(x,y,t)$ is the logarithm of the normalized density fluctuations, $\phi=\phi(x,y,t)$ is the normalized electrostatic potential, $B=B(x,y)$ is the normalized magnetic field, $T$ is the constant ion temperature normalized by the electron temperature and $\left[ f , g \right] = \mathbf{\hat{z}} \cdot \bm{\nabla}f \times \bm{\nabla}g$ is the canonical bracket in the plane across the magnetic field $\mathbf{B} = B \mathbf{\hat{z}}$. All the derivatives are defined on the perpendicular plane. The bracket $\left[ \phi , n \right]$ corresponds to the dynamics due to the $\mathbf{E} \times \mathbf{B}$ drift, and the bracket $\left[ n , 1/B \right]$ corresponds to the term driving the interchange instability.\\ 
When addressing the question of the Hamiltonian structure of the model~(\ref{eq:TOKAM:continuity})-(\ref{eq:TOKAM:vorticity}), it is easy to show that, provided that boundary terms vanish, a conserved quantity for the model is
\begin{eqnarray}
\displaystyle
I = \frac{1}{2}\int d^2x \Bigg( (1+T) n^{2} + \vert\bm{\nabla}\phi\vert^{2} \Bigg),
\label{eq:TOKAM:Inv}
\end{eqnarray}
where the first part corresponds to the potential energy for ions and electrons and the second part is the kinetic energy. This conserved quantity $I$ is interpreted as the total energy of the system and is a natural candidate for the Hamiltonian. Nevertheless, in spite of having a conserved quantity, the model (\ref{eq:TOKAM:continuity})-(\ref{eq:TOKAM:vorticity}) does not seem to possess a Hamiltonian formulation, at least of the Lie-Poisson type, which is the most common form for reduced fluid models. More precisely, Poisson brackets between functionals $F$ and $G$ of the form
\begin{eqnarray}  \label{brack}
\{ F , G \} = \sum_{i,j,k=1}^2 \int d^2x \Bigg( \left( V_{ij}^k \xi_i + \frac{\alpha_{j}^k}{B} \right) \left[ \frac{\delta F}{\delta \xi_j} , \frac{\delta G}{\delta \xi_k} \right] + \left( W_{ij}^k \xi_i + \frac{\alpha_{j}^k}{B} \right) \left[ \bm{\nabla} \frac{\delta F}{\delta \xi_j} ; \bm{\nabla} \frac{\delta G}{\delta \xi_k} \right] \Bigg),\nonumber
\end{eqnarray}
where the vector bracket is defined by $\left[ \mathbf{A} ; \mathbf{B} \right] = \sum_i \left[ A_i , B_i \right]$, the terms $\alpha_{j}^k$, $V_{ij}^k$ and $W_{ij}^k$ are constant, $\xi_1 = n$ , $\xi_2 = \Delta \phi$ and $\delta F/\delta \xi$ is the functional derivative of the functional $F$ with respect to $\xi$, fail to give the desired equations of motion~(\ref{eq:TOKAM:continuity})-(\ref{eq:TOKAM:vorticity}) when applied to $I$.
Even if a bracket giving Eqs.~(\ref{eq:TOKAM:continuity})-(\ref{eq:TOKAM:vorticity}) with the invariant $I$ defined by Eq.~(\ref{eq:TOKAM:Inv}) has been found, it does not satisfy Jacobi identity and thus is not a Poisson bracket.\\
The reason for the difficulty to find a Hamiltonian structure resides in the co-existence of compressibility terms (those containing $1/B$ in the continuity equation of the electrons) with the ion-gyroviscous term (in the vorticity equation). The compressibility terms are often retained in the continuity equation (see, e.g., Refs.~\cite{Haz85_86,Haz87,Eic96,Gar01}) in order to account for the fact that the velocity field is not incompressible in the presence of an inhomogeneous magnetic field. If these compressibility terms are neglected in the two-field model, the resulting system is Hamiltonian~\cite{Haz87,Dagnelund05}. On the other hand, it has been shown that the system obtained by keeping these terms and eliminating the ion-gyroviscous terms is also Hamiltonian~\cite{Haz87}. It is thus the simultaneous presence of the two contributions which seems to complicate the search for a Hamiltonian structure. This leads to the problem of finding a Hamiltonian model accounting for both compressibility terms and FLR corrections.\\
An elegant and effective way to introduce FLR corrections to a cold-ion model, while preserving the Hamiltonian structure, was introduced in Refs.~\cite{Morrison84,Haz87} and was referred to as the gyromap. The gyromap procedure was rigorously found from the Braginskii's closure for the stress tensor~\cite{Braginskii65} only for the three-field model described in Refs.~\cite{Hsu86,Haz87} which does not contain the compressibility terms in the continuity equation. In this paper, we apply the gyromap procedure to the cold-ion limit of the two-field model with the compressibility terms. We show that all the FLR correction terms produced by this method are obtained from the Braginskii's closure for the stress tensor, by means of an expansion based on a physically sound ordering.\\
The paper is organized as follows: In Sec.~2, we apply the gyromap procedure to the cold-ion version of the two-field model (\ref{eq:TOKAM:continuity})-(\ref{eq:TOKAM:vorticity}). In Sec.~3, we show that the resulting model is directly obtained from a systematic expansion.

\section{The gyromap}
The gyromap algorithm was introduced in Ref.~\cite{Haz87} in order to introduce FLR corrections to a cold-ion model, while preserving the Hamiltonian structure. The model with FLR corrections obtained through this procedure possesses the same bracket of the original cold-ion model, but different dynamical variables and Hamiltonian. The fact that the Poisson bracket is inherited from the cold-ion Hamiltonian model ensures that the resulting model is Hamiltonian. As a by-product, the Casimir invariants of the resulting model with FLR corrections become easily available. We also recall that, as mentioned in Ref.~\cite{Haz87}, the gyromap procedure possesses the desirable features that the cold-ion limit ($T=0$) of the post-gyromap model gives the initial cold-ion model, and that the diamagnetic effects predicted by the kinetic theory are conserved at the first order.
In what follows we use dimensionless quantities, i.e. we rescale space variables by the sonic Larmor radius $\rho_s$, the density by the equilibrium density $n_0$, the charge by the electric charge $e$, the electron and ion temperatures by the electron temperature $T_{\rm e}$, the magnetic field $B$ by the spatial mean value $B_0$ and time by the inverse of the ion cyclotron frequency $\omega_{c,{\rm i}}= e B_0/m_{\rm{i}}$ where $m_{\rm{i}}$ is the ion mass.\\
We start with the cold-ion version of Eqs.~(\ref{eq:TOKAM:continuity})-(\ref{eq:TOKAM:vorticity}) which describes the dynamical evolution of the plasma density $n$ and the vorticity $\Delta \phi$
\begin{eqnarray}
\displaystyle
\frac{\partial n}{\partial t} &=& - \left[ \phi , n \right] - \left[ \phi , \frac{1}{B} \right] + \left[ n , \frac{1}{B} \right], \label{eq:TOKAM-2D:n}\nonumber\\
\nonumber\\
\displaystyle
\frac{\partial \Delta\phi}{\partial t} &=& - \left[ \phi , \Delta\phi \right] + \left[ n , \frac{1}{B} \right].
\label{eq:TOKAM-2D:phi}\nonumber
\end{eqnarray}
We consider the system whose field variables are $n(x,y)$ and $\phi(x,y)$. Here we do not specify the time dependence of the variables which is implicitly assumed. In the algebra of observables which are functionals of $n$ and $\phi$, the Hamiltonian structure is defined by the Hamiltonian
\begin{eqnarray}
H(n,\phi) = \frac{1}{2}\int d^2x \Bigg( n^{2} + \vert\bm{\nabla}\phi\vert^{2} \Bigg),
\label{eq:TOKAM-2D:Ham}\nonumber
\end{eqnarray}
and by the Poisson bracket
\begin{eqnarray}
\displaystyle
\{ F , G \} = \int d^2x \Bigg( \left(n+\frac{1}{B}\right) \left( \left[ \frac{\delta F}{\delta n} , \frac{\delta G}{\delta n} \right] + \left[ \frac{\delta F}{\delta n} , \frac{\delta G}{\delta \Delta\phi} \right] + \left[ \frac{\delta F}{\delta \Delta\phi} , \frac{\delta G}{\delta n} \right] \right) + \Delta\phi \left[ \frac{\delta F}{\delta \Delta\phi} , \frac{\delta G}{\delta \Delta\phi} \right] \Bigg).
\label{eq:TOKAM-2D:Lie-Poisson}\nonumber
\end{eqnarray}
This bracket verifies all the required properties for a Poisson bracket like bilinearity, antisymmetry, Jacobi identity and Leibniz identity.\\
The gyromap procedure is initiated from the Hamiltonian defined by
\begin{eqnarray}
H(n,\Phi) = \frac{1}{2}\int d^2x \Bigg( (1+T)n^{2} + \vert\bm{\nabla}\Phi\vert^{2} \Bigg),
\label{eq:Gyromap:Ham}
\end{eqnarray}
where $\Phi$ is the stream function for the FLR model. The choice of Eq.~(\ref{eq:Gyromap:Ham}) as Hamiltonian is motivated by the requirement of adding the ion internal energy to the cold-ion Hamiltonian and having a kinetic energy term whose relation with the cold-ion kinetic energy is retrieved a posteriori. We introduce an auxiliary variable $\xi$ defined by $\Delta \Phi = \xi + T\Delta n/2$. The shift of $T\Delta n/2$ corresponds to a shift of half the magnetization velocity. This is the required transformation in order to yield the proper FLR corrections~\cite{Haz87}. The next step is to use the same Poisson bracket written in the new variables $(N,\xi)$,
\begin{eqnarray}
\displaystyle
\{ F , G \} = \int d^2x \Bigg( \left(N+\frac{1}{B}\right) \left( \left[ \frac{\delta F}{\delta N} , \frac{\delta G}{\delta N} \right] + \left[ \frac{\delta F}{\delta N} , \frac{\delta G}{\delta \xi} \right] + \left[ \frac{\delta F}{\delta \xi} , \frac{\delta G}{\delta N} \right] \right) + \xi \left[ \frac{\delta F}{\delta \xi} , \frac{\delta G}{\delta \xi} \right] \Bigg).
\label{eq:Gyromap:Lie-Poisson_N-xi}
\end{eqnarray}
The change of variables, $N=n$ and $\xi = \Delta \Phi - T \Delta n / 2$, includes a change of functional derivatives,
\begin{eqnarray}
\displaystyle
\frac{\delta}{\delta N} &=& \frac{\delta}{\delta n} + \frac{T}{2} \Delta \frac{\delta}{\delta \Delta \Phi},
\label{eq:Gyromap:fct_der_2}\nonumber\\
\nonumber\\
\displaystyle
\frac{\delta}{\delta \xi} &=& \frac{\delta}{\delta \Delta \Phi}. \label{eq:Gyromap:fct_der_1}\nonumber
\end{eqnarray}
Hence the bracket expressed in the variables $(n,\Phi)$, which is still a Poisson bracket, is the following,
\begin{eqnarray}
%\begin{array}{l}
\displaystyle
\{ F , G \} &=& \int d^2x \Bigg( \left(n+\frac{1}{B}\right) \left( \left[ \frac{\delta F}{\delta n} + \frac{T}{2}\Delta \frac{\delta F}{\delta \Delta \Phi} , \frac{\delta G}{\delta \Delta \Phi} \right] + \left[ \frac{\delta F}{\delta \Delta \Phi} , \frac{\delta G}{\delta n} + \frac{T}{2}\Delta \frac{\delta G}{\delta \Delta \Phi} \right] \right)\Bigg. \nonumber\\
\displaystyle
%\qquad\qquad
&&\Bigg. + \left(n+\frac{1}{B}\right) \left[ \frac{\delta F}{\delta n} + \frac{T}{2}\Delta \frac{\delta F}{\delta \Delta \Phi} , \frac{\delta G}{\delta n} + \frac{T}{2}\Delta \frac{\delta G}{\delta \Delta \Phi} \right] + \left( \Delta \Phi - \frac{T}{2} \Delta n \right) \left[ \frac{\delta F}{\delta \Delta \Phi} , \frac{\delta G}{\delta \Delta \Phi} \right] \Bigg).
%\end{array}
\label{eq:Gyromap:Lie-Poisson}
\end{eqnarray}
With the Hamiltonian defined by Eq.~(\ref{eq:Gyromap:Ham}), the equation of motion for the density is
\begin{eqnarray}
\displaystyle
\frac{\partial n}{\partial t} = \{ n , H \} = - \left[ \Phi + \frac{T}{2} \Delta \Phi , n + \frac{1}{B} \right] + (1+T) \left[ n , \frac{1}{B} \right]. \label{eq:Gyromap:continuity}
\end{eqnarray}
We assume that the continuity equation for electrons is unaffected by FLR corrections. Therefore comparing Eq.~(\ref{eq:TOKAM:continuity}) with Eq.~(\ref{eq:Gyromap:continuity}) gives the following relation between $\Phi$ and $\phi$:
\begin{eqnarray}
\displaystyle
\left( 1+\frac{T}{2}\Delta \right) \Phi = \phi + T n. \label{eq:Gyromap:Phi-vs-phi}
\end{eqnarray}
On the other hand, the equation for the generalized vorticity $\Delta\Phi$ is given by
\begin{eqnarray}
\displaystyle
\frac{\partial \Delta \Phi}{\partial t} = \{ \Delta \Phi , H \} = - \left[ \Phi , \Delta \Phi \right] + \left(1+T\right)\left(1+\frac{T}{2}\Delta\right)\left[ n , \frac{1}{B} \right] + T \bm{\nabla} \cdot \left[ n+\frac{1}{B} , \bm{\nabla} \Phi \right] - \frac{T^2}{4}\Delta \left[ \Delta \Phi , n+\frac{1}{B} \right].
\label{eq:Gyromap:vorticity}
\end{eqnarray}
Equations~(\ref{eq:Gyromap:continuity})-(\ref{eq:Gyromap:vorticity}) are the final result of the gyromap procedure. They include FLR corrections and the resulting system has a Hamiltonian structure since the bracket~(\ref{eq:Gyromap:Lie-Poisson}) is obtained from the Poisson bracket~(\ref{eq:Gyromap:Lie-Poisson_N-xi}) by a change of variables.\\
By expanding the operator  $(1+T\Delta/2)^{-1}$ for small $T$ and making use of Eq.~(\ref{eq:Gyromap:continuity}), it is possible to obtain an expression for Eq.~(\ref{eq:Gyromap:vorticity}) valid for small ion temperature $T$ (in terms of $n$ and $\phi$),
\begin{eqnarray}
\displaystyle
\frac{\partial \Delta\phi}{\partial t} &=& - \left[ \phi , \Delta\phi \right] + \left[ n , \frac{1}{B} \right]
+ T \left( \left[ n , \frac{1}{B} \right] - \bm{\nabla} \cdot \left[ n , \bm{\nabla}\phi \right] - \left[ \bm{\nabla} \phi \mathbf{;} \bm{\nabla} \Delta \phi \right] \right) \nonumber\\
\displaystyle
&&+ T^2 \left( \left[ \bm{\nabla}\Delta n \mathbf{;} \bm{\nabla}\phi \right] - \left[ \bm{\nabla} n \mathbf{;} \bm{\nabla}\Delta \phi \right] - 2 \left[ n , \Delta n \right] + 2 \Delta \left(1+\frac{\Delta}{2}\right) \left[ n , \frac{1}{B} \right] \right. \nonumber\\
\displaystyle
&&\left.
- \frac{\Delta}{4} \left[ \Delta \phi , n + \frac{1}{B} \right] + \frac{1}{2} \left[ \bm{\nabla}\phi \mathbf{;} \bm{\nabla}\Delta \phi \right] + \frac{1}{4} \left[ \Delta \phi , \Delta^2 \phi \right] \right) + O(T^3).
\label{eq:Gyromap:vorticity_2}
\end{eqnarray}
This expression is useful, because it allows one to directly compare the leading order terms of the vorticity equation~(\ref{eq:Gyromap:vorticity_2}), with the vorticity equation as given in Ref.~\cite{Dagnelund05}~[see also Eq.~(\ref{eq:TOKAM:vorticity})] which refers to a model where the only FLR term is $- T \bm{\nabla} \cdot \left[ n , \bm{\nabla}\phi \right]$. We notice that already at the order $O(T)$, the two equations differ from an FLR correction term given by $-T \left[ \bm{\nabla} \phi ; \bm{\nabla} \Delta \phi \right]$. From Eq.~(\ref{eq:Gyromap:vorticity_2}), it is also straightforward to see that, in the limit $T \rightarrow 0$, the FLR model is correctly reduces to the original cold-ion model, as expected.\\
In summary, in order to obtain a Hamiltonian structure with the ion temperature, the gyromap procedure starts from the Hamiltonian structure without ion temperature and then considers FLR effects using a change of variables.\\
In what follows, we consider Braginskii's closure for the stress tensor in order to derive a model with FLR terms and compare it with the above model given by Eqs.~(\ref{eq:Gyromap:continuity})-(\ref{eq:Gyromap:vorticity}).

\section{Derivation from Braginskii's closure}
We first consider the dynamics of a plasma composed of two species, ions and electrons. In a simplified magnetic geometry where the magnetic field is constant (but non-uniform) and its direction is fixed, we restrict the plasma dynamics to the plane transverse to the magnetic field lines, i.e. all dynamical variables only depend on the two coordinates $x$ and $y$. The dynamical variables are the ion and electron densities, $n_{\rm{i}}(x,y)$ and $n_{\rm{e}}(x,y)$, and the ion and electron velocity fields ${\bf v}_{\rm i}(x,y)$ and ${\bf v}_{\rm e}(x,y)$. We assume a quasi-neutrality hypothesis which allows us to reduce the number of variables, i.e. we consider that $n_{\rm i}(x,y)=n_{\rm e}(x,y)$ which is denoted $n(x,y)$ in what follows. The dynamical equation for the density is given by the continuity equation for the electrons,
\begin{equation}
\frac{\partial n}{\partial t} + \bm{\nabla} \cdot \left( n {\bf v}_{\rm e} \right) = 0.
\label{eq:conserv_density}\nonumber
\end{equation}
Due to the quasi-neutrality assumption, the same equation holds for ions, which translates into
\begin{equation}
\label{eq:qnj}
\bm{\nabla} \cdot \left( n({\bf v}_{\rm e}-{\bf v}_{\rm i})\right)=0.
\end{equation}
The dynamical equation for the velocity field ${\bf v}_s(x,y)$ of the species $s$, where $s$ refers to ions or electrons, is given by 
\begin{equation}
\mu_s n \left( \frac{\partial}{\partial t} + {\bf v}_s \cdot \bm{\nabla} \right) {\bf v}_s = n \left( {\bf E} + {\bf v}_s \times {\bf B} \right) - e_s T_s \bm{\nabla} n - e_s \bm{\nabla} \cdot \overline{\overline{{\bm \pi}_s}},
\label{eq:conserv_motion}
\end{equation}
where ${\bf E}$ is the electric field, $T_s$ is the dimensionless temperature of the species $s$ (i.e. $T_{\rm i}=T$ and $T_{\rm e}=1$), $\mu_s$ its dimensionless mass (i.e. $\mu_{\rm i}=1$ and $\mu_{\rm e} = m_{\rm e}/m_{\rm i} \ll \mu_{\rm i}$ because the mass of the electron $m_{\rm e}$ is negligible compared to the one of the ions $m_{\rm i}$) and $e_s$ its dimensionless charge (i.e. $e_{\rm i}=1$ and $e_{\rm e}=-1$). The stress tensor associated with the species $s$, denoted $\overline{\overline{{\bm \pi}_s}}$, is taken from Ref.~\cite{Braginskii65}. For each species, it is composed of viscosity terms identified by viscosity coefficients of various orders. Within a strong magnetic field approximation, we only consider the two higher order viscosity terms labeled by $\eta_0$ and $\eta_3$ with $\eta_0 \gg \eta_3$. The viscosity coefficients for ions are $\eta_0^{\rm{i}} = 0.96 n T \tau_{\rm i} \gg \eta_3^{\rm{i}} = n T/2$ where $\tau_s$ is the normalized collisional time for the species $s$.\\
In what follows, we only use the part of the stress tensor perpendicular to the magnetic field lines. Given these approximations, the divergence of the stress tensor for ions becomes
\begin{equation}
\bm{\nabla} \cdot \overline{\overline{{\bm \pi}_{\rm i}}} = - \alpha_{\rm i} \tau_{\rm i} T \bm{\nabla} \left( n \bm{\nabla} \cdot {\bf v}_{\rm i} \right) + \frac{T}{2} \left[ n\hz\times \Delta {\bf v}_{\rm i} - (\hz \times \bm{\nabla} n \cdot \bm{\nabla}) {\bf v}_{\rm i} + (\bm{\nabla} n\cdot \bm{\nabla})(\hz\times{\bf v}_{\rm i}) \right],\nonumber
\end{equation}
where $\alpha_{\rm{i}} = 0.32$. The viscosity coefficients for electrons are $\eta_0^{\rm{e}} = 0.73 n \tau_{\rm e} \gg \eta_3^{\rm{e}} = -n \mu_{\rm{e}}/2$ where $\tau_{\rm e}$ is the collisional time for the electrons. We consider the following approximation for the divergence of the stress tensor for the electrons
\begin{equation}
\bm{\nabla} \cdot \overline{\overline{{\bm \pi}_{\rm e}}} = - \alpha_{\rm e} \tau_{\rm e} \bm{\nabla} \left( n \bm{\nabla} \cdot {\bf v}_{\rm e}\right),\nonumber
\end{equation}
where $\alpha_{\rm{e}} = 0.24$. The dimensionless equation for the velocity field of each species becomes
\begin{eqnarray}
\mu_s n \left( \frac{\partial}{\partial t} + {\bf v}_s \cdot \bm{\nabla} \right) {\bf v}_s &=& - n \bm{\nabla} \phi + n B {\bf v}_s \times \hz - e_s T_s \bm{\nabla} n + \alpha_s\tau_s e_s T_s \bm{\nabla} \left( n\bm{\nabla}\cdot {\bf v}_s \right) \nonumber \\
&& - \mu_s \frac{e_s T_s}{2} \left[ n \hz \times \Delta {\bf v}_s - (\hz\times\bm{\nabla} n \cdot \bm{\nabla}) {\bf v}_s + (\bm{\nabla} n\cdot \bm{\nabla}) (\hz\times {\bf v}_s) \right].
\label{eq:conserv_motion_dimless}
\end{eqnarray}
We apply the cross product with $\mathbf{\hat{z}}/nB$ to Eq.~(\ref{eq:conserv_motion_dimless}) in order to obtain the velocity as the sum of the ${\bf E}\times {\bf B}$, the diamagnetic, the polarization drifts and some gyroviscous contributions from the stress tensor,
\begin{eqnarray}
{\bf v}_s &=& \frac{\hz \times \bm{\nabla} \left( \phi + e_sT_s \log  n \right) }{B} + \mu_s \frac{1}{B} \left( \frac{\partial}{\partial t} + {\bf v}_s \cdot \bm{\nabla} \right) \hz \times {\bf v}_s - \frac{\alpha_s\tau_s e_s T_s}{nB} \hz \times \bm{\nabla} ( n \bm{\nabla} \cdot {\bf v}_s ) \nonumber \\
&& - \mu_s \frac{e_s T_s}{2B}\left[ \Delta {\bf v}_s + (\hz\times \bm{\nabla} \log n \cdot \bm{\nabla}) (\hz\times {\bf v}_s) + (\bm{\nabla} \log n \cdot \bm{\nabla}) {\bf v}_s \right].\label{eq:velocity_s}
\end{eqnarray}
We expand all dimensionless quantities in the following way
\begin{eqnarray}
1/B &=& 1 + \varepsilon/\tilde{B}, \label{eq:ordering_B}\nonumber\\
\phi &=& \varepsilon \tilde{\phi}, \label{eq:ordering_phi}\nonumber\\
n &=& 1 + \varepsilon \tilde{n}, \label{eq:ordering_n}\nonumber\\
{\bf v}_s &=& \varepsilon {\bf v}_s^{(1)} + \varepsilon^2 {\bf v}_s^{(2)}. \label{eq:ordering_v}\nonumber
\end{eqnarray}
The small parameter $\varepsilon \ll 1$ denotes the amplitude of the fluctuations of the electromagnetic field as well as that of the dynamical variables $n$ and ${\bf v}_s$. Moreover, the temporal variations of these quantities are small so that the temporal variation is denoted $\varepsilon \partial/\partial t$. In what follows, we omit the tildas over the fluctuating quantities for simplicity.
Here we consider the model for weak collisionality (i.e.\ $\mu_s / \tau_s \ll 1$) such that $\tau_{\rm{i}}$ is of order $1/\varepsilon$ and $\tau_{\rm{e}} \ll 1/\varepsilon$. In addition $T$ is of order one, and the parameters $\alpha_s$ are of order $\varepsilon$ (i.e.\ $\alpha_{\rm{i}} \tau_{\rm{i}} T$ is of order $1$ and  $\alpha_{\rm{e}} \tau_{\rm{e}} \ll 1$). The leading order of Eq.~(\ref{eq:velocity_s}) (terms proportional to $\varepsilon$) gives
$$
\left( 1 + \mu_s \frac{e_s T_s}{2}\Delta \right) {\bf v}_s^{(1)} = \hz \times \bm{\nabla} \left( \phi + e_s T_s  n \right) - \alpha_s \tau_s e_s T_s \hz \times \bm{\nabla} ( \bm{\nabla} \cdot {\bf v}_s^{(1)} ).
$$
From the previous equation, we directly obtain that $\bm{\nabla} \cdot {\bf v}_{\rm e}^{(1)}=0$. From the expansion of Eq.~(\ref{eq:qnj}), we also conclude that $\bm{\nabla} \cdot {\bf v}_{\rm i}^{(1)}=0$. Therefore the equation for ${\bf v}_s^{(1)}$ becomes 
\begin{equation}
\label{eq:v1mu}
\left( 1 + \mu_s \frac{e_s T_s}{2}\Delta \right) {\bf v}_s^{(1)} = \hz \times \bm{\nabla} \left( \phi + e_s T_s  n \right),
\end{equation}
so that the first order velocity is the sum of the $\mathbf{E} \times \mathbf{B}$ velocity and the diamagnetic velocity with FLR corrections on the left hand side.  
The second order of the velocity ${\bf v}_s$ is given by
\begin{eqnarray}
\left( 1 + \mu_s \frac{e_s T_s}{2}\Delta \right) \mathbf{v}_s^{(2)} &=& \frac{1}{B}{\bf v}_s^{(1)} + \mu_s \left( \frac{\partial}{\partial t} + {\bf v}_s^{(1)} \cdot \bm{\nabla} \right) \hz \times {\bf v}_s^{(1)} - eT_s n \hz \times \bm{\nabla} n \nonumber\\
&&- \alpha_s \tau_s e_s T_s \hz \times \bm{\nabla} \left( \bm{\nabla} \cdot {\bf v}_s^{(2)} \right) - \frac{\mu_s e_s T_s}{2}\left[ (\hz\times\bm{\nabla} n\cdot\bm{\nabla}) (\hz\times {\bf v}_s^{(1)}) + (\bm{\nabla} n\cdot\bm{\nabla}) {\bf v}_s^{(1)} \right],
\label{eq:v2mu}
\end{eqnarray}
where we have used Eq.~(\ref{eq:v1mu}) for ${\bf v}_s^{(1)}$. 
We notice that we have also used the expansion $\bm{\nabla} \log(1+\varepsilon n)=\varepsilon \bm{\nabla} n - \varepsilon^2 n \bm{\nabla} n + O(\varepsilon^3)$.\\
As a first step, for the electrons, we see that from Eq.~(\ref{eq:v1mu}), the first order of the electron velocity is the sum of the $\mathbf{E} \times \mathbf{B}$ velocity and the electron diamagnetic velocity
\begin{equation}
{\bf v}_{\rm e}^{(1)} = \hz \times \bm{\nabla} \left( \phi - n \right),
\label{eq:ve1}
\end{equation}
and the second order of the electron velocity is
\begin{equation}
{\bf v}_{\rm e}^{(2)} = \frac{1}{B} {\bf v}_{\rm e}^{(1)} + n \hz \times \bm{\nabla} n + \alpha_{\rm e} \tau_{\rm e} \hz \times \bm{\nabla} \left( \bm{\nabla} \cdot {\bf v}_{\rm e}^{(2)} \right).
\label{eq:ve2}\nonumber
\end{equation}
From this equation, we get that the divergence of ${\bf v}_{\rm e}^{(2)}$ is equal to the divergence of ${\bf v}_{\rm e}^{(1)}/B$, and using Eq.~(\ref{eq:ve1}), it becomes
\begin{equation}
\bm{\nabla} \cdot {\bf v}_{\rm e}^{(2)} = \left[ \phi - n , \frac{1}{B} \right].
\label{eq:ve2_div}
\end{equation}
The continuity equation of the electron density at the second order in $\varepsilon$ becomes
$$
\frac{\partial n}{\partial t} + \bm{\nabla} \cdot {\bf v}_{\rm e}^{(2)} + {\bf v}_{\rm e}^{(1)} \cdot \bm{\nabla} n = 0,
$$
or equivalently, using Eqs.~(\ref{eq:ve1})-(\ref{eq:ve2_div}),
\begin{equation}
\frac{\partial n}{\partial t} = - \left[ \phi , n + \frac{1}{B} \right] + \left[ n , \frac{1}{B} \right].
\label{eq:Braginskii:continuity_e}\nonumber
\end{equation}
In fact, the first order of the ion velocity given by Eq.~(\ref{eq:v1mu}) suggests a change of variables $( 1+\frac{T}{2}\Delta ) \Phi = \phi + T n$ with which we can write the first order of the ion velocity into
\begin{equation}
{\bf v}_{\rm i}^{(1)} = \hz \times \bm{\nabla} \Phi.
\label{eq:vi1_1Phi}\nonumber
\end{equation}
We notice that this change of variables is also the one suggested by the gyromap [see Eq.~(\ref{eq:Gyromap:Phi-vs-phi})].
Using this variable $\Phi$, the continuity equation becomes
\begin{equation}
\label{eq:cee}
\frac{\partial n}{\partial t} = - \left[ \left( 1 + \frac{T}{2} \Delta \right) \Phi , n + \frac{1}{B} \right] + \left( 1 + T \right) \left[ n , \frac{1}{B} \right].
\end{equation}
As a second step, we work with the continuity equation for ions which is
\begin{equation}
\label{eq:cei}
\frac{\partial n}{\partial t} + \bm{\nabla} \cdot {\bf v}_{\rm i}^{(2)} + {\bf v}_{\rm i}^{(1)} \cdot \bm{\nabla} n = 0.
\end{equation}
First we obtain the following formula for the divergence of ${\bf v}^{(2)}_{\rm i}$ from Eq.~(\ref{eq:v2mu}),
\begin{equation}
\left( 1 + \frac{T}{2}\Delta \right) \bm{\nabla} \cdot {\bf v}_{\rm i}^{(2)} = - \frac{\partial \Delta \Phi}{\partial t} - \left[ \Phi , \Delta \Phi \right] + \left[ \Phi , \frac{1}{B} \right] - \frac{T}{2} \left[ \Delta \Phi , n \right] - T \left[ \bm{\nabla} \Phi ; \bm{\nabla} n \right].
\label{eq:vi2_div}
\end{equation}
Next we multiply Eq.~(\ref{eq:cei}) by $\left(1+T \Delta / 2 \right)$ in order to use the Eq.~({\ref{eq:vi2_div}) and we insert Eq.~(\ref{eq:cee}) so as to obtain
\begin{equation}
\frac{\partial \Delta \Phi}{\partial t} = - \left[ \Phi , \Delta \Phi \right] + \left( 1 + T \right) \left( 1 + \frac{T}{2}\Delta \right) \left[ n , \frac{1}{B} \right] + T \bm{\nabla} \cdot \left[ n + \frac{1}{B} , \bm{\nabla} \Phi \right] - \frac{T^2}{4} \Delta \left[ \Delta \Phi , n + \frac{1}{B} \right].
\label{eq:Braginskii:vorticity}
\end{equation}
We notice that Eqs.~(\ref{eq:cee})-(\ref{eq:Braginskii:vorticity}) coincide with Eqs.~(\ref{eq:Gyromap:continuity})-(\ref{eq:Gyromap:vorticity}), which were obtained by applying the gyromap. Therefore, we have shown that the terms generated from the gyromap are obtained consistently from Braginskii's closure for the stress tensor by making use of an appropriate ordering.\\
\\
In summary, it is possible to construct a model with FLR corrections and its Hamiltonian structure from a cold-ion model which possesses a Hamiltonian structure by applying a gyromap procedure which generates all the relevant FLR terms at the leading order.
The change of variables introduced by the gyromap is directly given by the definition of the stream function at the first order of the ion velocity. We have shown that the two-field reduced model~(\ref{eq:Gyromap:continuity})-(\ref{eq:Gyromap:vorticity}) is obtained using the Braginskii's closure for the stress tensor by considering an apt ordering on the dynamical variables.

\section*{Acknowledgements}
We acknowledge financial support from the Agence Nationale de la Recherche. 
This work was supported by the European Community under the contract of Association between EURATOM, CEA and the French Research Federation for fusion studies. The views and opinions expressed herein do not necessarily reflect those of the European Commission.


\begin{thebibliography}{AAA00}
\expandafter\ifx\csname url\endcsname\relax\def\url#1{\texttt{#1}}\fi
\bibitem{Morrison80PLA}
{P.~J.~Morrison},
{\it}{ Phys. Lett. A} \textbf{80}, 383 (1980).
\bibitem{Marsden82}
{J.~E.~Marsden and A.~Weinstein},
{\it}{ Physica D} \textbf{4}, 058102 (1982).
\bibitem{Morrison80PRL}
{P.~J.~Morrison and J.~M.~Greene},
{\it}{ Phys. Rev. Lett.} \textbf{45}, 790 (1980).
\bibitem{Morrison05}
{P.~J.~Morrison},
{\it}{ Phys. Plasmas} \textbf{12}, 058102 (2005).
\bibitem{Grasso01}
{D.~Grasso, F.~Califano, F.~Pegoraro and F.~Porcelli},
{\it}{ Phys. Rev. Lett.} \textbf{86}, 5051 (2001).
\bibitem{Tassi10}
{E.~Tassi, P.~J.~Morrison, D.~Grasso and F.~Pegoraro},
{\it}{ Nucl. Fusion} \textbf{50}, 034007 (2010).
\bibitem{Haz85_86}
{R.~D.~Hazeltine, M.~Kotschenreuther and P.~J.~Morrison},
{\it}{ Phys. Fluids} \textbf{28}, 2466 (1985);{\it}{ Phys. Fluids} \textbf{29}, 341 (1986).
\bibitem{Hsu86}
{C.~T.~Hsu, R.~D.~Hazeltine and P.~J.~Morrison},
{\it}{ Phys. Fluids} \textbf{29}, 1480 (1986).
\bibitem{Brizard92}
{A.~J.~Brizard},
{\it}{ Phys. Fluids B} \textbf{4}, 1213 (1992).
\bibitem{Mikhailovskii71}
{A.~B.~Mikhailovskii and V.~S.~Tsypin},
{\it}{ Plasma Phys.} \textbf{13}, 785 (1971).
\bibitem{Hinton71}
{F.~L.~Hinton and C.~W.~Horton},
{\it}{ Phys. Fluids} \textbf{14}, 116 (1971).
\bibitem{Scott03}
{B.~Scott},
{\it}{ Phys. Plasmas} \textbf{10}, 963 (2003).
%\bibitem{Haz85_86}
%{R.~D.~Hazeltine, M.~Kotschenreuther and P.~J.~Morrison},
%{\it}{ Phys. Fluids} \textbf{29}, 341 (1986).
\bibitem{Haz87}
{R.~D.~Hazeltine, C.~T.~Hsu and P.~J.~Morrison},
{\it}{ Phys. Fluids} \textbf{30}, 3204 (1987).
\bibitem{Eic96}
{T.~Eickermann and K.~H.~Spatschek},
{\it}{ Phys. Plasmas} \textbf{3}, 2869 (1996). 
\bibitem{Gar01}
{O.~E.~Garcia},
{\it}{ J. Plasma Phys.} \textbf{65}, 81 (2001). 
\bibitem{Dagnelund05}
{D.~Dagnelund and V.~P.~Pavlenko},
{\it}{ Phys. Scr.} \textbf{71}, 293 (2005). 
\bibitem{Morrison84}
{P.~J.~Morrison, I.~L.~Caldas and H.~Tasso},
{\it}{ Z. Naturfors.  Sect. A-J. Phys. Sci.} \textbf{39}, 1023 (1984).
\bibitem{Braginskii65}
{S.~I.~Braginskii},
{\it}{ Rev. Plasma Phys.} \textbf{1}, 205 (1965).
\end{thebibliography}
\end{document}